\documentclass[11pt]{article}
\usepackage{amsmath}
\usepackage[paperwidth=90mm]{geometry}
\usepackage{bbm}
\usepackage{hyperref}
\hypersetup{colorlinks=true, linkcolor=blue, citecolor=blue}
\usepackage{anysize}
\usepackage{bm}
\usepackage{natbib}
\usepackage{xcolor}
\usepackage{amsthm}
\usepackage{amssymb}
\usepackage{graphicx} 
\usepackage{multirow}
\usepackage{tabularx}
\usepackage{adjustbox}
\usepackage{bibunits}
\usepackage[normalem]{ulem}
\usepackage{titling}
\usepackage{authblk}
\marginsize{1in}{1in}{1in}{1in}
\newtheorem{theorem}{Theorem}

\newtheorem{remark}{Remark}
\newtheorem{proposition}{Proposition}
\newtheorem{assumption}{Assumption}
\newtheorem{corollary}{Corollary}

\usepackage{adjustbox}
\usepackage{caption}
\usepackage{xcolor}
\usepackage[normalem]{ulem}
\usepackage{multirow}
\usepackage{threeparttable}
\newcommand{\amin}{\operatornamewithlimits{arg\,min}}

\def\indep{\bot\!\!\!\bot}

\def\ap{a^\prime}

\def\zd{\mathrm{d}}
\def\lV{\Vert}
\def\rV{\Vert}

\def\sp{\text{sp}}

\def\T{{ \mathrm{\scriptscriptstyle T} }}

\def\calL{{\mathcal{L}}}
\allowdisplaybreaks[4]

\begin{document}
	\title{Proximal Mediation Analysis with Unmeasured Treatment-Induced Confounding}
    \author[1]{Xiaoying Zhang}
	\author[2]{Jiawei Shan$^*$}
	\author[1]{Wei Li\thanks{Co-corresponding authors. E-mails: \texttt{jiawei.shan@wisc.edu} and \texttt{weilistat@ruc.edu.cn}.}}
	\affil[1]{Center for Applied Statistics and School of Statistics, Renmin University of China}
	\affil[2]{Department of Biostatistics \& Medical Informatics, University of Wisconsin-Madison}
    \begin{NoHyper}
    \maketitle
    \end{NoHyper}

    \begin{abstract}
    Mediation analysis provides a central framework for elucidating causal mechanisms, yet its application is often impeded by treatment-induced confounding, under which the widely used natural mediation effects are generally unidentifiable. Interventional effects have been proposed as an alternative when these confounders are observable; however, identifying and estimating interventional effects remains challenging when  confounders are unmeasured. In this paper, we address this issue by using observed variables as proxies for unmeasured treatment-induced confounders. We establish four proximal identification results and develop a  multiply robust, semiparametric locally efficient estimator that accommodates flexible machine learning methods for nuisance parameter estimation. The proposed approach is illustrated through simulation studies and a real-data application evaluating racial disparities in life satisfaction mediated by discrimination.
    \end{abstract}

\vfill

\newpage

\section{Introduction}
Understanding how and why a treatment affects an outcome is an essential question in empirical research. Mediation analysis provides a powerful framework to uncover these mechanisms and has been widely applied across  disciplines such as epidemiology, psychology, and economics. In recent decades, causal mediation analysis has advanced substantially, particularly through  nonparametric frameworks that decompose the total effect into the natural direct effect (NDE) and the natural indirect effect (NIE) \citep{robins1992identifiability,pearl2001direct,imai2010general,tchetgen2012semiparametric}. Identification of natural effects is most commonly based on the assumption that all confounders are pre-treatment \citep{ImaiKeeleYamamoto2010}. This assumption, however, is violated in the presence of treatment-induced confounders. 
Such confounders are common in practice, particularly when treatment is assigned well before the mediator is measured.
For example, \cite{miles2017quantifying} investigated the path-specific effects of  treatment regimen assignment on virologic failure through adherence over the second six months, in which drug toxicity and adherence over the first six months constitute treatment-induced confounders. In our motivating application, we study racial disparities in life satisfaction mediated by perceived discrimination, where socioeconomic status is a mediator-outcome confounder that may itself be affected by race. In addition, such confounders are often measured only imperfectly, which further complicates both identification and inference.

Even when all confounders are observed, treatment-induced confounding poses unique challenges for mediation analysis, and several methods have been developed for this case. For the NDE and NIE, point identification can be achieved under additional assumptions such as monotonicity \citep{tchetgen2014identification,rudolph2023efficient} or no treatment heterogeneity \citep{xia2023identification}. 
A separate line of work reconceptualizes direct and indirect effects. \citet{avin2005identifiability} introduced path-specific effects to capture the contributions of distinct causal pathways and showed that certain of these effects are identifiable under suitable conditions. Building on this idea, \citet{miles2020semiparametric}  developed a semiparametric  framework for efficient inference of path-specific effects.
Interventional effects, outlined by \citet{petersen2006estimation, vanderlaan2008direct} and further developed by \citet{vanderweele2014effect}, define direct and indirect effects by replacing the potential mediator with a random draw from its observed distribution. This framework has been extended to accommodate time-varying treatments and mediators \citep{vanderweele2017mediation}, and  \citet{diaz2021nonparametric} further proposed nonparametric efficient estimators for interventional effects. 
Despite these advances, all of these methods require treatment-induced confounders to be fully observed. When some are unmeasured, failing to adjust for them can lead to biased estimates, and no existing method addresses this case. 

In the presence of unmeasured pre-treatment confounding, the recently developed proximal causal inference framework provides a principled strategy for mitigating confounding bias in observational studies \citep{miao2018identifying,shi2020multiply,tchetgen2024introduction,cui2024semiparametric}. By leveraging  proxy variables for unmeasured confounders, such as noisy measurements or negative controls  that satisfy specific independence and relevance conditions, it identifies causal effects that would otherwise be unidentifiable. This framework has recently been extended to causal mediation analysis.  \cite{dukes2023proximal} proposed a proximal  approach for identifying NDE and NIE under unmeasured confounding. \cite{bai2025proximal} developed a proximal method for the population intervention indirect effect (PIIE), a measure designed for mediation analysis with harmful treatments that differs from the interventional indirect effect central to our work. \cite{ghassami2025causal} further studied settings with an unobserved mediator, showing that the NDE and NIE are identifiable when no unmeasured confounding is present, while the PIIE remains identifiable even under unmeasured treatment-outcome confounding. These developments, however, all address unmeasured pre-treatment confounding and do not apply to the unmeasured treatment-induced confounders that are the focus of this paper.

In this paper, we develop a proximal mediation analysis approach under unmeasured treatment-induced confounding, an important yet previously unexplored setting. To our knowledge, this is the first work to extend proximal causal inference from pre-treatment to treatment-induced unmeasured confounding for identifying interventional direct and indirect effects.
Our contributions are threefold.
First, we establish nonparametric identification of the interventional direct and indirect effects from a pair of proxy variables for the unmeasured treatment-induced confounder, and propose four distinct identification strategies that rely on different combinations of nuisance functions.
Second, we derive the efficient influence function (EIF) for the causal parameters of interest and develop a multiply robust estimator that is consistent and asymptotically normal whenever at least one of the four sets of nuisance models is correctly specified, and that achieves the semiparametric efficiency bound when all working models are correct. 
Third, building on the EIF, we propose a debiased machine learning estimator that estimates the bridge functions via minimax learning and the remaining nuisance functions via flexible machine learning, retaining root-$n$ consistency and asymptotic normality under slower nuisance rates. 
Our approach accommodates both continuous and binary mediators, and is illustrated through simulation studies and an empirical application evaluating racial disparities in life satisfaction mediated by discrimination.

The remainder of the paper is organized as follows. Section~\ref{sec:identification} reviews identification under measured treatment-induced confounding and develops our proximal identification for the unmeasured case. Section~\ref{sec:estimation} establishes the semiparametric theory and presents the multiply robust estimators, leveraging both parametric and debiased machine learning methods. Section~\ref{sec:simulation} evaluates the finite-sample performance of the proposed estimators through simulation studies. Section~\ref{sec:application} presents an empirical application evaluating racial disparities in life satisfaction mediated by discrimination. Section~\ref{sec:discussion} concludes. All proofs are given in the supplementary material.

\section{Nonparametric identification}
\label{sec:identification}
\subsection{Preliminaries}
\label{subsec:preliminaries}

We consider a setting where the goal is to assess the effect of a binary treatment $A\in \{0,1\}$ on an outcome $Y$ that is mediated via a variable $M$. Let $L$ and $U$ denote the observed and unobserved treatment-induced confounders, respectively, which are intermediate variables affected by the treatment and in turn confound the mediator-outcome relationship. The observed pre-treatment or baseline covariates are denoted as $X$. Let $Y(a,m)$ denote the potential outcome that would be observed if the treatment and mediator variables were set to values $A=a$ and $M=m$ (possibly contrary to fact). Similarly, let $M(a)$ denote the potential value of the mediator had the treatment variable been set to the value $A=a$. We use $f(\cdot)$ to denote a generic probability density or mass function and assume that $M$, $L$, $U$, and $X$ take values in $\mathcal{M}$, $\mathcal{L}$, $\mathcal{U}$, and $\mathcal{X}$, respectively. We first posit the following standard assumptions that are commonly adopted in causal inference.
\begin{assumption}[Consistency]
\label{ass-consistency}
    (i) $M(a)=M$  if $A=a$; (ii) $Y(a,m)=Y$  if $A=a$ and $M=m$.
\end{assumption}
\begin{assumption}[Positivity]
\label{ass-positivity}
    (i) $f(a\mid X)>0$ almost surely for $a\in\{0,1\}$; (ii) $f(u,l\mid A,X)>0$ almost surely for  $u\in \mathcal{U},l\in \mathcal{L}$; (iii) $f(m\mid A,U,L,X)>0$ almost surely for  $m\in \mathcal{M}$.
\end{assumption}

{The presence of treatment-induced confounders violates the cross-world independence assumption that underlines the natural effects \citep{imai2010general,ImaiKeeleYamamoto2010}. As a result, the NDE and NIE, defined respectively as ${E}\left\{Y(1,M(0))-Y(0,M(0))\right\}$ and ${E}\left\{Y(1,M(1))-Y(1,M(0))\right\}$, are generally not identifiable from observed data, whether or not the confounders are measured. In fact, even when no such confounders exist, cross-world independence remains a strong and empirically untestable assumption, undermining the credibility of identified natural effects.}
To address this challenge, we build on the seminal work on interventional effects \citep{petersen2006estimation,vanderlaan2008direct,vanderweele2014effect}, defining direct and indirect effects through stochastic interventions on the mediator while preserving the  decomposition structure of the natural effects. Let $G(a)$ be a random draw from the distribution of $M(a)$ given $X$. Interventional effects replace the natural mediator value $M(a)$ in these contrasts with the stochastic draw $G(a)$, and the interventional direct and indirect effects are defined as follows:
\begin{equation}
\label{decomposition}
    \underbrace{{E}\left\{Y(1,G(1))-Y(0,G(0))\right\}}_{\text{Total effect}}=\underbrace{{E}\left\{Y(1,G(1))-Y(1,G(0))\right\}}_{\text{Interventional indirect effect}}+\underbrace{{E}\left\{Y(1,G(0))-Y(0,G(0))\right\}}_{\text{Interventional direct effect}}.
\end{equation}
The interventional direct effect captures the effects of causal pathways not operating through the mediator, while the interventional indirect effect quantifies the effect through pathways involving the mediator. {However, unlike natural effects, the interventional direct and indirect effects do not sum to the average treatment effect, and they admit a distinct causal interpretation. In our application, for instance, the interventional indirect effect measures how life satisfaction would change if the distribution of perceived discrimination were shifted from what it would be were everyone Black or African American to what it would be were everyone White or Caucasian. This effect is therefore more policy-relevant than the corresponding natural effect, as it conrresponds to a concrete intervention on the discrimination distribution, particularly when the treatment itself, here race, cannot be directly manipulated.} We refer readers to \cite{vanderweele2014effect}, \cite{vansteelandt2017interventional}, and \cite{diaz2021nonparametric} for further discussion of these distinctions. 
Path-specific effects \citep{avin2005identifiability,miles2020semiparametric} are another alternative to the natural effects under treatment-induced confounding, but their definition unavoidably invokes the potential values $U(a)$ of the confounder, which makes them hard to interpret when the confounder is abstract and unmeasured.
By contrast, the interventional effect in \eqref{decomposition} are defined only through the distribution of the potential mediator, so they remain interpretable even when the treatment-induced confounding is unmeasured.

In what follows, we focus on the identification and estimation of {$\psi^{a,\ap}={E}\{Y(a,G(\ap))\}$} for any $a,\ap\in \{0,1\}$, from which the effects decomposed in \eqref{decomposition} can be obtained. 
We further make the following latent conditional ignorability assumption.
\begin{assumption}[Latent conditional ignorability]
\label{ass-ignorability}
    (i) $Y(a,m)\indep A\mid X$; (ii) $M(a)\indep A\mid X$; (iii) $Y(a,m)\indep M\mid A,U,L,X$.
\end{assumption}
Assumption \ref{ass-ignorability} essentially requires that $X$ captures all potential confounding sources of the treatment-outcome and treatment-mediator relationships, and $A,U,L,X$ captures all potential confounding sources of the mediator-outcome relationship. Under Assumptions~\ref{ass-consistency}-\ref{ass-ignorability}, \cite{vanderweele2014effect} showed that $\psi^{a,\ap}$ can be expressed as
\begin{equation}
\label{identification_prev}
    \psi^{a,\ap}=\iiiint {E}(Y\mid a,u,l,m,x)f(u,l\mid a,x)f(m\mid \ap,x)f(x)\zd u\zd l\zd m\zd x.
\end{equation}
In the special case where $U=\emptyset$, 
Assumption~\ref{ass-ignorability}(iii) reduces to $Y(a,m)\indep M\mid A,L,X$, which posits no unmeasured confounding of the $M$-$Y$ relationship. Accordingly, equation~\eqref{identification_prev} reduces to the identification formula presented in \citet{vanderweele2014effect}. 
	
\subsection{Proximal identification}
\label{subsec:proxiamal_identification}
In many practical scenarios, some treatment-induced confounders cannot be directly observed, and the identification formula \eqref{identification_prev} is no longer feasible. 
In this section, we consider the setting where $U\neq\emptyset$ and establish the identification of the parameters of interest, assuming access to two proxy variables $Z$ and $W$ that serve as imperfect representations of the unmeasured treatment-induced confounder $U$. 
Specifically, let $Z$ be the \emph{mediator-inducing} proxy, which is a potential cause of $M$ and, after controlling for observed variables, relates to $Y$ only through the unmeasured confounder $U$; similarly, let $W$ be the \emph{outcome-inducing} proxy, which is a potential cause of $Y$ and, after controlling for observed variables, relates to $M$ only through $U$.
The following assumption  formally specifies the conditions for valid proxies.
\begin{assumption}[Proxies]
\label{ass-proxies}
    (i) $Z\indep Y\mid A,U,L,M,X$; (ii) $W\indep (M,Z)\mid A,U,L,X$.
\end{assumption}

\begin{figure}
    \centering
    \includegraphics[width=0.4\linewidth]{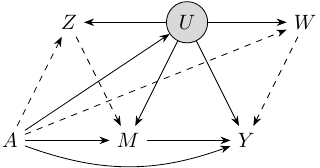}
    \caption{The causal diagram with unmeasured treatment-induced confounder $U$, with $L$ and $X$ omitted for simplicity.}
    \label{fig-unobserved}
\end{figure}

Assumption \ref{ass-proxies} essentially requires no directed edges from $Z$ to $Y$, and from $M$ to $W$. Also, $Z$ and $W$ are only associated via their common measured and unmeasured causes $A,L,X$ and $U$. Figure \ref{fig-unobserved} displays a graphical illustration of the proxy variables that satisfy these conditional independence conditions, where the dashed edges can be present or absent. Actually, there are many other diagrams that may be compatible with Assumption \ref{ass-proxies}. For example, the potential edge from $Z$ to $M$ can be flipped, or there exists an extra bi-directed edge between $U$ and $W$ or $U$ and $Z$, etc. This provides flexibility for researchers when choosing variables as proxy variables.
It is also worth noting that compared with \cite{dukes2023proximal}, which exploits proxy variables to address the unmeasured pre-treatment confounding of $A$, $M$, and $Y$, our requirements for proxies are weaker since we allow for direct edges between $A$ and $W$, and between $Z$ and $M$. An intuition behind this is that instead of being a common cause of $A$, $M$, and $Y$, the variable $U$ only confounds the $M$-$Y$ relationship in our setting, where $A$ acts as an observable confounder between $M$ and $Y$ that is not causally affected by the unmeasured confounder $U$. {In this sense, if we treat $A$ as a covariate and $M$ as the treatment, then Figure \ref{fig-unobserved} resembles the diagram used in conventional proximal causal inference for estimating the average treatment effect. 
Our target estimands, however, are the interventional direct and indirect effects, which are functionals of the potential-mediator distribution rather than a single average effect of $M$ on $Y$. Consequently, conventional proximal methods cannot be directly applied to identify these estimands or to account for the treatment-induced confounder $U$, and we develop dedicated proximal identification results for this purpose in the remainder of this section.
}
We further impose the following conditions formalizing an informational relevance requirement that the proxies must fulfill.
\begin{assumption}[Completeness]
\label{ass-completeness}
    ~\newline(i) For any square-integrable function $g(u)$, if ${E}\{g(U)\mid Z,A=a,L=l,M=m,X=x\}=0$ almost surely for any $l$, $m$, and $x$, then $g(U)=0$ almost surely.
    \newline(ii) For any square-integrable function $g(u)$, if ${E}\{g(U)\mid W,A=a,L=l,M=m,X=x\}=0$ almost surely for any $l$, $m$, and $x$, then $g(U)=0$ almost surely.
\end{assumption}
These conditions are formally known as completeness conditions which can accommodate both categorical and continuous confounders. Intuitively, Assumption \ref{ass-completeness} requires that the set of proxies must have sufficient variability relative to the variability of the unmeasured confounder $U$. In the case of categorical $Z$, $W$, and $U$ with the number of categories $d_{Z}$, $d_{W}$, and $d_{U}$, respectively, the assumption requires $d_{Z}\geq d_{U}$ and $d_{W}\geq d_{U}$; that is, $Z$ and $W$ must each have at least as many categories as $U$. For continuously distributed confounders, many commonly used parametric and semiparametric models, such as exponential families \citep{newey2003instrumental}, satisfy the completeness condition.

Now we can give our first identification result based on the so-called outcome-confounding bridge function.
\begin{theorem}
\label{theorem-identify1}
    Suppose that there exists an outcome confounding bridge function $h_{a}(w,l,m,x)$ that satisfies
    \begin{equation}
    \label{bridge_outcome_Z}
        {E}(Y\mid Z,A=a,L,M,X)={E}\{h_a(W,L,M,X) \mid Z,A=a,L,M,X\}.
    \end{equation}
    Then, under Assumptions \ref{ass-consistency}-\ref{ass-proxies} and \ref{ass-completeness}(i), it follows that
    \begin{align*}
    \label{bridge_outcome_U}
        {E}(Y\mid U,A=a,L,M,X)={E}\{h_a(W,L,M,X) \mid U,A=a,L,M,X\},
    \end{align*}
    and furthermore that $\psi^{a,\ap}$ is identified by $\psi^{a,\ap}=\psi_1$, where
    \begin{align*}
    \psi_1={E}\{\tau_{a,\ap}(X)\},
    \end{align*}
    and $\tau_{a,\ap}(X)=\iiint h_a(w,l,m,X)f(w,l\mid a,X)f(m\mid \ap,X)\zd w\zd l\zd m.$
\end{theorem}
Equation (\ref{bridge_outcome_Z}) defines an inverse problem known as a Fredholm integral equation of the first kind \citep{kress1989linear}, and formal technical conditions for the existence of a solution to similar equations can be found in \cite{miao2018identifying}, \cite{li2023non}, and \cite{cui2024semiparametric}. {In particular, the following completeness condition, along with the regularity conditions provided in Section S2 of the supplementary material, suffices for the existence of a solution to (\ref{bridge_outcome_Z}). 
\begin{assumption}
    For any square-integrable function $g(z)$, if ${E}\{g(Z)\mid W,A=a,L=l,M=m,X=x\}=0$ almost surely for any $l$, $m$, and $x$, then $g(Z)=0$ almost surely.
\end{assumption}}

While Theorem \ref{theorem-identify1} does not require the uniqueness of a solution to the integral equation (\ref{bridge_outcome_Z}), all solutions lead to a unique value of $\psi^{a,\ap}$. 
The quantity $\psi_1$ represents a fully marginalized version of the target estimand. Specifically, $\psi_1$ is obtained by integrating $h_a(W,L,M,X)$ with respect to the conditional distribution of $(W,L)$ given $(a,X)$, the conditional distribution of $M$ given $(a',X)$, and then taking expectation over the marginal distribution of $X$.
It has a similar form with the proximal g-formula proposed by \cite{tchetgen2024introduction} and the proximal mediation formula proposed by \cite{dukes2023proximal}. Two additional representations for the identification formula, which follow immediately from algebraic manipulation of $\psi_1$, are presented in the following corollary.  
\begin{corollary}
\label{corollary1}
    The identification formula $\psi_1$ in Theorem \ref{theorem-identify1} can be equivalently expressed as  the following two forms:\\
    (i) \[\psi_1={E}\left\{\frac{\mathbb{I}(A=\ap)}{f(\ap\mid X)}\gamma_a(M,X)\right\},\] where $\gamma_a(M,X)=\iint h_a(w,l,M,X)f(w,l\mid a,X)\zd w\zd l$, and $\mathbb{I}(\cdot)$ is an indicator function;\\
    (ii) \[\psi_1={E}\left\{\frac{\mathbb{I}(A=a)}{f(a\mid X)}\eta_{a,\ap}(W,L,X)\right\},\] where $\eta_{a,\ap}(W,L,X)=\int h_a(W,L,m,X)f(m\mid \ap,X)\zd m$.
\end{corollary}
For clarity, we denote the expressions in (i) and (ii) by $\psi_2$ and $\psi_3$, respectively. 
Both $\psi_2$ and $\psi_3$ represent partially marginalized and partially weighted versions of the target estimand. They are defined as inverse probability weighted averages of conditional expectations of $h_a(W,L,M,X)$,  where $\psi_2$ corresponds to the conditional expectation over $(W,L)$ given $(a,X)$, and $\psi_3$ corresponds to the conditional expectation over $M$ given $(a',X)$.
In the following, we further introduce a mediation confounding bridge function, based on which we establish an alternative proximal identification result.
\begin{theorem}
\label{theorem-identify2}
    Suppose that there exists a mediation confounding bridge function $q_{a}(z,l,m,x)$ that satisfies
    \begin{equation}
    \label{bridge_mediator_W}
        \frac{1}{f(M\mid W,A=a,L,X)}={E}\{q_a(Z,L,M,X) \mid W,A=a,L,M,X\}.
    \end{equation}
    Then, under Assumptions \ref{ass-consistency}-\ref{ass-proxies} and \ref{ass-completeness}(ii), it follows that
    \begin{align*}
    \label{bridge_mediator_U}
    \frac{1}{f(M\mid U,A=a,L,X)}={E}\{q_a(Z,L,M,X) \mid U,A=a,L,M,X\},
    \end{align*}
    and furthermore that $\psi^{a,\ap}$ is identified by $\psi^{a,\ap}=\psi_4$, where
    \begin{align*}
        \psi_4={E}\left\{\frac{\mathbb{I}(A=a)}{f(a\mid X)}{f(M\mid \ap,X)}q_a(Z,L,M,X)Y\right\}.
    \end{align*}
\end{theorem}
Similar to (\ref{bridge_outcome_Z}), (\ref{bridge_mediator_W}) also defines a Fredholm integral equation of the first kind. {Analogously, the following completeness condition, along with the regularity conditions  similar to those described above, suffices for the existence of a solution to (\ref{bridge_mediator_W}). 
\begin{assumption}
    For any square-integrable function $g(w)$, if ${E}\{g(W)\mid Z,A=a,L=l,M=m,X=x\}=0$ almost surely for any $l$, $m$, and $x$, then $g(W)=0$ almost surely.
\end{assumption}}
The quantity $\psi_4$ is a fully weighted version of the target estimand, where the weight is given by the product of the inverse probability weight, the conditional density of $M$ given $(\ap,X)$, and the mediation confounding bridge function. 

We therefore have four results for proximal identification, each involving a distinct set of observed data distributions. It is of interest to consider how our results degenerate to the case where all confounders are observable. Actually, suppose that $U=\emptyset$, then under Assumption \ref{ass-proxies}, the equation (\ref{bridge_outcome_Z}) reduces to
\[{E}(Y\mid A=a,L,M,X)={E}\{h_a(W,L,M,X) \mid A=a,L,M,X\},\]
which leads to the solution $h_a(W,L,M,X)={E}(Y\mid W,A=a,L,M,X)$. 
Also, equation (\ref{bridge_mediator_W}) reduces to 
\[\frac{1}{f(M\mid A=a,L,X)}={E}\{q_a(Z,L,M,X) \mid A=a,L,M,X\},\]
which leads to the solution $q_a(Z,L,M,X)={1}/{f(M\mid Z,A=a,L,X)}$.
By substituting the two solutions into the expressions of $\psi_j,j=1,\ldots,4$, we derive identification results under no unmeasured treatment-induced confounding.

\section{Estimation}
\label{sec:estimation}

\subsection{Semiparametric theory}
The identification functionals presented in the last section can be directly used to construct plug-in estimators for the parameter of interest. 
However, such estimators will typically inherit first-order bias from the estimation of nuisance functions and are sensitive to model misspecification. In this section, we instead turn to semiparametric efficiency theory and employ the efficient influence function (EIF) to develop a locally efficient and multiply robust estimator, which can provide protection against certain types of model misspecification while maintaining consistency. 

Under Assumptions \ref{ass-consistency}-\ref{ass-completeness}, the four identification formulas presented in Section \ref{subsec:proxiamal_identification} are equivalent, and we denote their common value by $\psi$. We consider inference for $\psi$ under the semiparametric model $\mathcal{M}_{\sp}$ that places no restrictions on the observed data distribution other than the existence (but not necessarily uniqueness) of a bridge function $h_a(W,L,M,X)$ that solves (\ref{bridge_outcome_Z}). The assumed existence of the outcome bridge functions places restrictions on the tangent space, and an additional regularity condition is imposed to obtain the EIF under $\mathcal{M}_{\sp}$. For a generic random variable (vector) $T$, we denote the space of all square-integrable functions of $T$ by $\calL_2(T)$. Let $\mathcal{S}_a:\calL_2(W,L,M,X) \rightarrow \calL_2(Z,L,M,X)$ be the conditional expectation operator given  by $(\mathcal{S}_ag)(Z,L,M,X) = {E}\left\{g(W,L,M,X)\mid Z,A=a,L,M,X\right\}$.
\begin{assumption}
\label{ass-regularity}
     {For any $a\in\{0,1\}$}, the operator $\mathcal{S}_a$ is surjective.
\end{assumption}
Surjectivity of $\mathcal{S}_a$ implies that for any function $f\in \calL_2(Z,L,M,X)$, there exists a function $g\in \calL_2(W,L,M,X)$, such that $\mathcal{S}_a$ maps $g$ to $f$. In practical terms, this can be realized if $W$ and $Z$ are both high-quality measurements of the underlying treatment-induced confounder $U$, with minimal and non-discrepant noise, such that variation in $Z$ is recoverable from $W$ via the conditional expectation map. 

In what follows, we use  $h_a$ and $q_a$ as shorthand for $h_a(W,L,M,X)$ and $q_a(Z,L,M,X)$, respectively, whenever no confusion arises. Let $\Delta=\{h_a,q_a,f(A\mid X),f(M\mid A,X),f(W,L\mid A,X)\}$ denote the set of all nuisance functions, and let ${O}=(X,Z,W,A,L,M,Y)$ denote the observed variables. Then we have the following result.
\begin{theorem}
\label{theorem-EIF}
    Under the semiparametric model $\mathcal{M}_{\sp}$ satisfying (\ref{bridge_outcome_Z}), the efficient influence function of $\psi$ evaluated at a law where (\ref{bridge_mediator_W}) and Assumption \ref{ass-regularity} hold, and $h_a$ and $q_a$ are uniquely defined, is given by $\mathrm{EIF}(O)=N(O;\Delta)-\psi$, where
    \begin{align*}
        \begin{aligned}
            &N(O;\Delta)=\frac{\mathbb{I}(A=a)}{f(a\mid X)}{f(M\mid \ap,X)}q_a(Z,L,M,X)\{Y-h_a(W,L,M,X)\}\\
            &~~~~~~~~~~~~~~+\frac{\mathbb{I}(A=a)}{f(a\mid X)}\left\{\eta_{a,\ap}(W,L,X)-\tau_{a,\ap}(X)\right\}+\frac{\mathbb{I}(A=\ap)}{f(\ap\mid X)}\left\{\gamma_a(M,X)-\tau_{a,\ap}(X)\right\}+\tau_{a,\ap}(X).
        \end{aligned}
    \end{align*}
    Therefore, the corresponding semiparametric local efficiency bound of $\psi$ equals ${E}\left\{\mathrm{EIF}^2(O)\right\}$.
\end{theorem}
As noted in \cite{ying2023proximal} and \cite{ghassami2025causal}, while Assumption \ref{ass-regularity} and the uniqueness of $h_a$ and $q_a$ are strong conditions, they are only needed for local efficiency statements and are not actually used for identification, estimation, or inference. 
In other words, even if these assumptions do not hold,  estimators introduced below remain regular and asymptotically linear under standard regularity conditions, though possibly at some loss of efficiency.

\subsection{Multiply robust estimation}
The EIF presented in Theorem \ref{theorem-EIF} implies that $\psi={E}\{N(O;\Delta)\}$, which motivates us to adopt the estimator $\hat{\psi}={P}_n\{N(O;\hat{\Delta})\}$, where ${P}_n(\cdot)$ represents the sample averaging operator, and $\hat{\Delta}=\{\hat{h}_a,\hat{q}_a,\hat{f}(A\mid X),\hat{f}(M\mid A,X),\hat{f}(W,L\mid A,X)\}$ denotes an estimator of $\Delta$, which may be derived from parametric models with finite-dimensional parameters, flexible nonparametric approaches, or machine learning methods. 

We first consider estimating the nuisance functions in $\Delta$ using a parametric model-based approach. Noting that the conditional densities $f(A\mid X),f(M\mid A,X),f(W,L\mid A,X)$ can be conveniently estimated via maximum likelihood under parametric models, our focus is on the estimation of $h_a$ and $q_a$. Let $h_a(W,L,M,X;\beta_a)$ and $q_a(Z,L,M,X;\theta_a)$ denote the parametric models for $h_a$ and $q_a$, indexed by finite-dimensional parameters $\beta_a$ and $\theta_a$.
Since $h_a(W,L,M,X)$ solves the integral equation (\ref{bridge_outcome_Z}), an estimator $\hat{\beta}_a$ can be obtained by solving the corresponding estimating equation:
\begin{align*}
    {P}_{n}\left[\mathbb{I}(A=a)\{Y-h_a(W,L,M,X;\beta_a)\}d(Z,L,M,X)\right]=0,
\end{align*}
where $d(Z,L,M,X)$ is a function of the same dimension as $\beta_a$. 
In contrast, estimating $q_a$ is relatively more challenging as the integral equation \eqref{bridge_mediator_W} involves an additional nuisance function $f(M\mid W,A=a,L,X)$. A further complication is that the nuisance components $f(W,L\mid A,X)$, $f(M\mid A,X)$, and $f (M\mid W,A,L,X)$ are not variationally independent, and therefore, introducing a separate parametric specification for $f(M\mid W,A,L,X)$ may induce model incompatibility. 
Previous studies have proposed using a copula to reparameterize the joint distribution (e.g., $f(M,W,L\mid A,X)$) in order to address similar challenges.
In this paper, however, we avoid introducing extra nuisance parameters and circumvent this problem based on the following Proposition \ref{proposition1}, which provides an equivalent moment equation of~(\ref{bridge_mediator_W}) that $q_a$ satisfies. 
Let $\mu$ denote a $\sigma$-finite dominating measure on $\mathcal{M}$. For a given function $\pi:\mathcal{M}\times\mathcal{X}\rightarrow\mathbb{R}$, assume that {$\pi(m,x)$} is measurable and uniformly square-integrable, i.e., $\sup_{x\in\mathcal{X}}\int\pi^2(m,x)\zd \mu (m)<\infty.$ Let $\mathcal{T}_\pi:\calL_2(W,L,M,X)\rightarrow \calL_2(W,L,X)$ be an operator given by \[(\mathcal{T}_\pi g)(W,L,X)=\int g(W,L,m,X)\pi(m,X)\zd \mu(m).\]  
\begin{proposition}
    \label{proposition1}
    The function $q_a(z,l,m,x)$ solves the integral equation (\ref{bridge_mediator_W}) if and only if it solves the following integral equation:
    \begin{equation}
    \label{equation-estimation_q}
        {E}\left[\mathbb{I}(A=a)\left\{\pi(M,X)q_a(Z,L,M,X)g(W,L,M,X)-(\mathcal{T}_\pi g)(W,L,X)\right\}\right]=0,
    \end{equation}
    where $g(W,L,M,X)$ is an arbitrary square-integrable function.
\end{proposition}
Note that although equation~(\ref{bridge_mediator_W}) involves the conditional density $f(M\mid W,A=a,L,X)$, the equivalent formulation in~(\ref{equation-estimation_q}) removes this dependence. Proposition \ref{proposition1} is important because directly solving~(\ref{bridge_mediator_W}) would require estimating $f(M\mid W,A=a,L,X)$, which can be difficult when $M$ is continuous or high-dimensional. In contrast, equation~(\ref{equation-estimation_q}) provides an alternative representation that avoids the need to estimate this conditional density, thereby facilitating the  estimation of $q_a(Z,L,M,X)$ through an appropriate choice of the function $\pi(m,x)$.

Specifically, let $\pi(m,x)=(-1)^{(1-m)}$ when $M$ is binary, (\ref{equation-estimation_q}) reduces to
\begin{align*}
    {E}\left[\mathbb{I}(A=a)\{(-1)^{(1-M)}q_a(Z,L,M,X)g(W,L,M,X)-g(W,L,1,X)+g(W,L,0,X)\}\right]=0,
\end{align*}
which motivates us to adopt the following estimating equation for $\theta_{a}$:
\begin{align*}
    {P}_{n}\left[\mathbb{I}(A=a)\{(-1)^{(1-M)}q_a(Z,L,M,X;\theta_{a})g(W,L,M,X)-g(W,L,1,X)+g(W,L,0,X)\}\right]=0,
\end{align*} 
where $g(W,L,M,X)$ is a function of the same dimension as $\theta_{a}$. 

For continuous $M$ and fixed $a^\prime$, let $\pi(m,x)=f(m\mid a^\prime,x)$ denote the conditional density of $m$ given $(a^\prime,x)$, then the moment condition (\ref{equation-estimation_q}) suggests the following estimating equation for $\theta_{a}$: 
\begin{align*}
    {P}_{n}\left[\mathbb{I}(A=a)\left\{f(M\mid a^\prime,X)q_a(Z,L,M,X;\theta_a)g(W,L,M,X)-(\mathcal{T}_{\pi,n}g)(W,L,X)\right\}\right]=0,
\end{align*} 
where $\mathcal{T}_{\pi,n}$ denotes the empirical version of the operator $\mathcal{T}_\pi$. Specifically, $\mathcal{T}_{\pi,n}g$ is obtained by taking the estimated conditional expectation of $g(W,L,M,X)$ over $M$ given $(a',X)$. Hence, estimating $q_a$ requires modeling the conditional distribution of $M$ given $(A,X)$. It is worth noting that $f(M\mid A,X)$ is already a nuisance function appearing in the EIF, so the proposed estimation procedure does not introduce any additional nuisance parameters.
\begin{remark}
In practice, the parametric forms of $h_a$ and $q_a$ depend on the true data-generating mechanisms of $Y$ and $M$, and it is easy to specify parametric models for them in some simple settings. For example, if $Y$ follows a linear model, or if $M$ follows a logit model of other variables, then $h_a$ is a linear function of $W,L,M,X$, and $q_a$ is a transformed version of a linear function of $Z,L,M,X$. In such cases, the corresponding choices for $d(Z,L,M,X)$ and $g(W,L,M,X)$ are naturally $(1,Z,L,M,X)$ and $(1,W,L,M,X)$, respectively.
\end{remark}

Once the nuisance functions in $\Delta$ have been estimated, one can construct the parametric model-based multiply robust estimator, denoted as $\hat{\psi}_{\text{mr}}$. The following theorem formalizes the multiple robustness of $\hat{\psi}_{{\text{mr}}}$, and we say bridge functions $h_a$ and $q_a$ are correctly specified if they satisfy equations (\ref{bridge_outcome_Z}) and (\ref{bridge_mediator_W}), respectively. 

\begin{theorem}
    Under {standard regularity conditions}, $\hat{\psi}_{\text{\normalfont mr}}$ is a consistent and asymptotically normal estimator of $\psi$ under the model $\mathcal{M}_{\text{\normalfont union}}=\mathcal{M}_1\cup \mathcal{M}_2\cup \mathcal{M}_3\cup \mathcal{M}_4$, where 
    \begin{align*}
        &\mathcal{M}_1: h_a(W,L,M,X),f(W,L\mid A,X) \text{~and~} f(M\mid A,X) \text{~are correctly specified}.\\
        &\mathcal{M}_2: h_a(W,L,M,X),f(W,L\mid A,X) \text{~and~} f(A\mid X)\text{~are correctly specified}.\\
        &\mathcal{M}_3: h_a(W,L,M,X),f(M\mid A,X) \text{~and~} f(A\mid X)\text{~are correctly specified}.\\
        &\mathcal{M}_4: q_a(Z,L,M,X),f(M\mid A,X) \text{~and~} f(A\mid X)\text{~are correctly specified}.
    \end{align*}
    Furthermore, $\hat{\psi}_{\text{\normalfont mr}}$ attains the semiparametric efficiency bound at the intersection submodel $\mathcal{M}_1\cap \mathcal{M}_2\cap \mathcal{M}_3\cap \mathcal{M}_4$ if Assumption \ref{ass-regularity} also holds.
\end{theorem}

Under particular specifications of the nuisance functions in $\mathcal{M}_1$, $\mathcal{M}_2$, $\mathcal{M}_3$, and $\mathcal{M}_4$, 
we can construct four different estimators $\hat{\psi}_1$, $\hat{\psi}_2$, $\hat{\psi}_3$, and $\hat{\psi}_4$, corresponding to the proximal identification results in Section~\ref{subsec:proxiamal_identification}. Under standard regularity conditions, $\hat{\psi}_j$ is consistent and asymptotically normal when model $\mathcal{M}_j$ holds, $j=1,\dots,4$. In contrast, the multiply robust estimator $\hat{\psi}_{\text{mr}}$ remains consistent and asymptotically normal provided that at least one of the four models is correct, thereby improving robustness to model misspecification.

\subsection{Debiased machine learning}
\label{subsec:DML}
Although our proposed parametric model-based multiply robust estimator mitigates sensitivity to model misspecification, practical implementations may still exhibit misspecification bias, especially in high-dimensional or complex settings. 
The efficient influence function we derived in Theorem~\ref{theorem-EIF} satisfies Neyman orthogonality, which naturally permits a more flexible double/debiased machine learning approach that accommodates modern, data-adaptive methods for nuisance estimation while preserving $\sqrt{n}$-consistency and asymptotic normality of the target estimator under mild regularity conditions \citep{chernozhukov2018double,kennedy2024semiparametric}.

A complication arises in our setting because $h_a$ and $q_a$ are defined as solutions to integral equations, for which conventional regression-based estimation techniques are not directly applicable. Recent studies by \cite{kallus2021causal} and \cite{ghassami2022minimax} have developed minimax learning procedures for  proximal causal inference to address similar challenges, drawing on the nonparametric adversarial learning framework of \cite{dikkala2020minimax}. Building on this line of research, we adapt and tailor a minimax learning approach to estimate the bridge functions $h_a$ and $q_a$ within our framework. Let $\mathcal{H}$ and $\mathcal{Q}$ denote normed function spaces equipped with norms $\|\cdot\|_{\mathcal{H}}$ and $\|\cdot\|_{\mathcal{Q}}$, respectively. Based on the moment conditions \eqref{bridge_outcome_Z} and \eqref{equation-estimation_q}, we formulate the following regularized optimization-based estimators for $h_a$ and $q_a$:
\begin{align*}
    \hat{h}_{a}(W,L,M,X)=\amin_{h\in\mathcal{H}}\sup_{q\in\mathcal{Q}}{P}_{n}\big[&\mathbb{I}(A=a)\{(Y-h(W,L,M,X))q(Z,L,M,X)\\
    &-q^2(Z,L,M,X)\}\big]-\lambda_{\mathcal{Q}}^h\|q\|_{\mathcal{Q}}^2+\lambda_{\mathcal{H}}^h\|h\|_{\mathcal{H}}^2,
\end{align*}
\begin{align*}
    \hat{q}_a(Z,L,M,X)=\amin_{q\in\mathcal{Q}}\sup_{h\in\mathcal{H}}{P}_{n}\big[&\mathbb{I}(A=a)\{\pi(M,X)q(Z,L,M,X)h(W,L,M,X)\\
    &-(\mathcal{T}_{\pi,n}h)(W,L,X)-h^2(W,L,M,X)\}\big]-\lambda_{\mathcal{H}}^q\|h\|_{\mathcal{H}}^2+\lambda_{\mathcal{Q}}^q\|q\|_{\mathcal{Q}}^2,
\end{align*}
where 
$\lambda_{\mathcal{Q}}^h,\lambda_{\mathcal{H}}^h,\lambda_{\mathcal{H}}^q,\lambda_{\mathcal{Q}}^q>0$ are regularized parameters.
The proposed minimax procedure accommodates various function classes such as reproducing kernel Hilbert spaces (RKHS) and neural networks. In our simulation studies, we employ the RKHS for illustration due to its computational convenience, and we refer readers to \cite{ghassami2022minimax} for closed-form solutions and convergence analysis of the corresponding estimators. 

Given the EIF and estimators for the nuisance functions, we estimate the target parameters using cross-fitting, which helps mitigate the need for restrictive Donsker conditions \citep{schick1986asymptotically,zheng2011cross,chernozhukov2018double}. Given a fixed positive integer $K$, we first divide the entire sample $\{1,\dots, n\}$ into $K$ disjoint folds $\{\mathcal{I}_1,\dots,\mathcal{I}_K\}$, with sample sizes $\{n_1,\dots,n_K\}$. Let $\mathcal{I}_{-k}=\{1,\dots, n\}\backslash\mathcal{I}_k$ denote the complement set of the $k$th partition.
For $k\in \{1,\dots,K\}$, we train the nuisance models using the data with index set $\mathcal{I}_{-k}$, and apply the estimated nuisances only to the hold-out fold indexed by $\mathcal{I}_{k}$. Let 
$\hat{\Delta}^{(k)}=\{\hat{h}_a^{(k)},\hat{q}_a^{(k)},\hat{f}^{(k)}(A\mid X),\hat{f}^{(k)}(M\mid A,X),\hat{f}^{(k)}(W,L\mid A,X)\}$ denote the machine learning estimators for nuisance models obtained from $\mathcal{I}_{-k}$. The minimax optimization technique described above can be used for obtaining $\hat{h}_a^{(k)}$ and $\hat{q}_a^{(k)}$, and standard regression techniques, in conjunction with a kernel density estimation procedure, can be applied to obtain $\hat{f}^{(k)}(A\mid X)$, $\hat{f}^{(k)}(M\mid A,X)$, and $\hat{f}^{(k)}(W,L\mid A,X)$. Finally, the debiased machine learning estimator of $\psi$ is given by:
\[\hat{\psi}_{\text{dml}}=\frac{1}{K}\sum_{k=1}^K{P}_{n_k}\{N(O;\hat{\Delta}^{(k)})\},\]
where ${P}_{n_k}$ denotes the empirical mean in the $k$th partition. 
Let $\|g\|=\{\int g^2(o)\zd F(o)\}^{1/2}$ denote the $\calL_2$ norm of any function $g$, and $\mathcal{S}_a':\calL_2(Z,L,M,X) \rightarrow \calL_2(W,L,M,X)$ be the adjoint operator of $\mathcal{S}_a$.
In the following theorem, we further establish the asymptotic properties of the debiased machine learning estimator $\hat{\psi}_{\text{dml}}$, provided that weak convergence rate requirements for the working models are met.
\begin{theorem}
\label{theorem-converge}
    Assume that for each $k$, the following conditions hold:\\
    \noindent (i) $\hat\Delta^{(k)}$ converges to $\Delta$ in probability;\\
    \noindent (ii) $\{\lvert\hat{h}_a^{(k)}\lvert,\lvert\hat{q}_a^{(k)}\rvert\}<C$, $\epsilon<\{f(A\mid X),\hat{f}^{(k)}(A\mid X)\}<1-\epsilon$, and $\epsilon<\{f(M\mid A,X),\hat{f}^{(k)}(M\mid A,X),\hat{f}^{(k)}(W,L\mid A,X)\}<C$ for some $0<\epsilon<1/2$ and $C>0$;\\
    (iii)  $n^{1/2}$-convergence of second-order terms, i.e., 
    \begin{align*}
        &\lV h_a-\hat{h}_a^{(k)}\rV
        \Big\{\lV \mathcal{S}_a'(q_a-\hat{q}_a^{(k)})\rV+\lV{f(A\mid X)}-{\hat{f}^{(k)}(A\mid X)}\rV+\lV f(M\mid A,X)-\hat{f}^{(k)}(M\mid A,X)\rV\Big\}\\
        +&\Vert f(W,L\mid A,X)-\hat{f}^{(k)}(W,L\mid A,X)\Vert\Big\{\Vert f(M\mid A,X)-\hat{f}^{(k)}(M\mid A,X)\Vert+\Vert {f(A\mid X)}-{\hat{f}^{(k)}(A\mid X)}\Vert\Big\}\\
        +&\lV{f(A\mid X)}-{\hat{f}^{(k)}(A\mid X)}\rV\lV f(M\mid A,X)-\hat{f}^{(k)}(M\mid A,X)\rV=o_{{p}}(n^{-1/2}).
    \end{align*}
    Then $\hat{\psi}_{\text{\normalfont dml}}$ is an asymptotically normal and semiparametrically efficient estimator.
\end{theorem}
Conditions (i)-(iii) in Theorem \ref{theorem-converge} are similar to those for debiased machine learning estimation of average
treatment effects \citep{kennedy2017non,chernozhukov2018double,kennedy2024semiparametric}. 
In particular, condition (iii) permits slower convergence rates for the nuisance function estimates and can be fulfilled when all the estimators converge at a rate faster than $n^{-1/4}$. Under appropriate structural assumptions, many highly data-adaptive machine learning techniques, such as RKHS and neural networks, can meet this requirement.

\section{Simulation}
\label{sec:simulation}
In this section, we conduct simulation studies to evaluate the finite-sample performance of the proposed estimators for $\psi^{1,0}$. For parametric estimators, we generate the baseline covariate $X$ from the normal distribution $N(0,1)$. The binary treatment variable $A$, treatment-induced confounders $L$ and $U$, and proxies $W$ and $Z$ are generated separately according to the following mechanisms: 
\begin{align*}
    A\mid X &\sim \text{Bernoulli}\left\{\text{expit}(X)\right\},\\
    L\mid A,X &\sim N(0.2-0.25A+0.3X,0.25^2),\\
    U\mid A,L,X &\sim N(0.5-0.3A+0.7L+0.8X,0.5^2),\\
    W\mid U,L,X &\sim N(0.5+0.5U-0.6L-0.15X,0.25^2),\\
    Z\mid A,U,L,X &\sim N(-0.3+0.4A+0.5U-0.6L+0.5X,0.25^2),
\end{align*}
where $\text{expit}(x) = \text{exp}(x)/\{1 + \text{exp}(x)\}$.
We assume that $U$ is an unmeasured variable and consider two scenarios in which the mediator $M$ is either binary or continuous. Specifically, $M$ is generated from a Bernoulli distribution in the binary case and from a normal distribution in the continuous case:
\begin{align*}
    \text{Binary case}~~
    M\mid A,U,L,X &\sim \text{Bernoulli}\left\{\text{expit}(-0.6+0.7A+0.5U+0.25L+X)\right\},\\ 
    \text{Continuous case}~~
    M\mid A,U,L,X &\sim N(-0.6+0.7A+0.5U+0.25L-0.4X,1).
\end{align*}
Finally, the outcome $Y$ is generated from the normal distribution:
\begin{align*}
    Y\mid A,U,L,M,X &\sim N(0.2+0.5A+U-0.3L+0.8M-0.6X,0.5^2).
\end{align*}
Since the conditional mean of $W$ is a linear function of $U$, $L$, and $X$, the data-generating mechanism is compatible with the outcome confounding bridge function model $h_a(W,L,M,X;\beta_a)=(1,W,L,M,X)^\T\beta_{a}$. We also show in Section S3.1 of the supplementary material that the compatible choices of mediation confounding bridge functions are $q_a(Z,L,M,X;\theta_{a})=1+\text{exp}\{(-1)^{1-M}(1,Z,L,M,X)^\T\theta_{a}\}$
for binary $M$, and 
$q_a(Z,L,M,X;\theta_{a})=1+\text{exp}\{(1,\mathfrak{X})^\T\theta_{a}\}$
for continuous $M$, where $\mathfrak{X}$ is a vector encompassing all linear terms, quadratic terms, and pairwise interaction terms of $Z,L,M,X$.

\begin{table}[]
\caption{Bias, standard error (SE), and 95\% coverage probability (CP) of the parametric estimators under $n = 1000$. The bias is reported in units of $10^{-2}$.}
\label{tabel1}
\resizebox{\textwidth}{!}{
\begin{tabular}{cccccccccccccccccccc}
\hline
\multicolumn{1}{l}{} & \multicolumn{3}{c}{Scenario 0} &  & \multicolumn{3}{c}{Scenario 1} &  & \multicolumn{3}{c}{Scenario 2} &  & \multicolumn{3}{c}{Scenario 3} &  & \multicolumn{3}{c}{Scenario 4} \\ \cline{2-4} \cline{6-8} \cline{10-12} \cline{14-16} \cline{18-20} 
 & Bias & SE & CP &  & Bias & SE & CP &  & Bias & SE & CP &  & Bias & SE & CP &  & Bias & SE & CP \\ \hline
 & \multicolumn{19}{c}{Binary $M$} \\
$\hat{\psi}_{\text{mr}}$ & -0.13 & 0.05 & 0.96 &  & -0.05 & 0.05 & 0.97 &  & 0.06 & 0.06 & 0.95 &  & -0.06 & 0.06 & 0.96 &  & -0.11 & 0.05 & 0.96 \\
$\hat{\psi}_1$ & -0.15 & 0.04 & 0.94 &  & -0.15 & 0.04 & 0.94 &  & -9.26 & 0.04 & 0.45 &  & 25.49 & 0.05 & 0.00 &  & 13.51 & 0.04 & 0.09 \\
$\hat{\psi}_2$ & -0.24 & 0.05 & 0.95 &  & -22.52 & 0.05 & 0.01 &  & -0.24 & 0.05 & 0.95 &  & 25.41 & 0.06 & 0.01 &  & 13.42 & 0.05 & 0.26 \\
$\hat{\psi}_3$ & -0.12 & 0.04 & 0.96 &  & 22.13 & 0.04 & 0.00 &  & -9.24 & 0.05 & 0.46 &  & -0.12 & 0.04 & 0.95 &  & -2.84 & 0.04 & 0.90 \\
$\hat{\psi}_4$ & -0.59 & 0.05 & 0.96 &  & 24.12 & 0.06 & 0.01 &  & -8.61 & 0.06 & 0.68 &  & 1.74 & 0.06 & 0.95 &  & -0.59 & 0.05 & 0.95 \\
 & \multicolumn{19}{c}{Continuous $M$} \\
$\hat{\psi}_{\text{mr}}$ & 0.11 & 0.07 & 0.96 &  & 0.02 & 0.06 & 0.96 &  & -0.02 & 0.07 & 0.96 &  & 0.30 & 0.07 & 0.97 &  & 0.23 & 0.07 & 0.96 \\
$\hat{\psi}_1$ & -0.20 & 0.06 & 0.95 &  & -0.20 & 0.06 & 0.95 &  & -5.11 & 0.06 & 0.85 &  & 24.65 & 0.07 & 0.06 &  & 10.67 & 0.06 & 0.57 \\
$\hat{\psi}_2$ & -0.38 & 0.06 & 0.96 &  & -15.90 & 0.06 & 0.24 &  & -0.38 & 0.06 & 0.96 &  & 24.60 & 0.07 & 0.07 &  & 10.54 & 0.06 & 0.61 \\
$\hat{\psi}_3$ & -0.19 & 0.06 & 0.96 &  & 17.96 & 0.07 & 0.26 &  & -5.10 & 0.06 & 0.86 &  & -0.19 & 0.06 & 0.96 &  & -4.06 & 0.06 & 0.90 \\
$\hat{\psi}_4$ & 0.24 & 0.08 & 0.96 &  & 16.55 & 0.07 & 0.39 &  & -2.64 & 0.07 & 0.95 &  & 2.72 & 0.07 & 0.95 &  & 0.24 & 0.08 & 0.96 \\
\cline{1-20}
\end{tabular}}
\end{table}
\begin{figure}[htbp]
    \centering
    \includegraphics[width=1\linewidth]{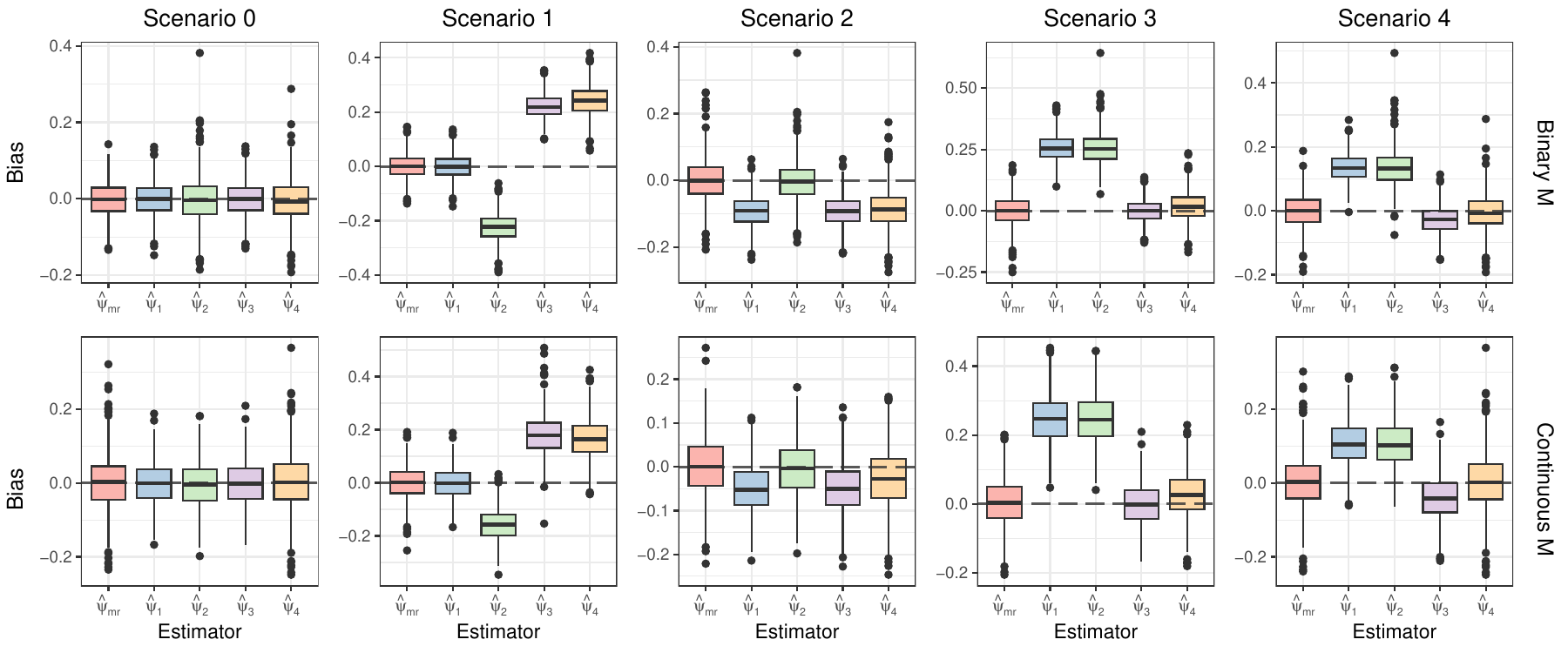}
    \caption{Box plots of bias of parametric estimators. First row corresponds to the case where $M$ is binary, second row corresponds to the case where $M$ is continuous. Columns from left to right correspond to Scenario 0-4, respectively.}
    \label{fig3-sim_par}
\end{figure}
We evaluate the performance of the proposed multiply robust estimator $\hat{\psi}_{\text{mr}}$, along with four alternative estimators, $\hat{\psi}_1$, $\hat{\psi}_2$, $\hat{\psi}_3$, and $\hat{\psi}_4$ under five scenarios. In Scenario 0, all nuisance functions are correctly specified, whereas in Scenarios 1-4, only the models in $\mathcal{M}_j,~j=1,\dots,4$ are correctly specified. Model misspecification is induced by replacing $X$ in the nuisance functions with $\lvert X\rvert^{1/2}$ in the binary $M$ case and $\lvert -X^2 + X + 1\rvert$ in the continuous $M$ case.
Table \ref{tabel1} summarizes the simulation results for all estimators under the five scenarios with a sample size of $n = 1000$, including the bias, standard error, and 95\% coverage probability, each averaged over 1000 replications.
Figure \ref{fig3-sim_par} presents the corresponding box plots of the bias.
As expected, the proposed estimator $\hat{\psi}_{\text{mr}}$ shows small bias and achieves nominal coverage probability across all scenarios, even under varying model misspecifications, thereby confirming its multiple robustness property. In contrast, the other four parametric estimators tend to exhibit larger bias and lower coverage probabilities when their corresponding nuisance models are misspecified. 
\begin{table}[]
\centering
\caption{Bias, standard error (SE), and 95\% coverage probability (CP) of $\hat{\psi}_{\text{\normalfont dml}}$ under $n\in\{250,500,1000\}$.  The bias is reported in units of $10^{-2}$.}
\label{tabel2}
\begin{tabular}{cccccccc}
\hline
 & \multicolumn{3}{c}{Binary $M$} &  & \multicolumn{3}{c}{Continuous $M$} \\ \cline{2-4} \cline{6-8} 
$n$  & Bias & SE & CP &  & Bias & SE & CP \\ \hline
250 & -3.95 & 0.16 & 0.94 &  & -7.17 & 0.21 & 0.94 \\
500 & -1.31 & 0.10 & 0.94 &  & -3.56 & 0.12 & 0.96 \\
1000 & -0.47 & 0.07 & 0.95 &  & -2.21 & 0.08 & 0.94 \\ \hline
\end{tabular}
\end{table}

{To evaluate the performance of the proposed debiased machine learning estimator $\hat{\psi}_{\text{dml}}$, we adopt a more complex data-generating mechanism that involves nonlinearity, interactions, and skewed distributions. Specifically, we set the observed $X$ and $L$ as the cubes of their true values, introduce interaction terms when generating $M$ and $Y$, and draw $W$ and $Z$ from log-normal distributions. The full data-generating process is detailed in Section S3.2 of the supplementary material.}
We consider a five-fold cross-fitting with $h_a$ and $q_a$ estimated by the method proposed in Section \ref{subsec:DML}. Neural networks \citep{ripley1996pattern} are employed to estimate the other nuisance functions, with conditional densities of continuous variables derived by fitting a Gaussian kernel to the estimated residuals. The estimation of $\eta_{a,\ap}(W,L,X)$, $\gamma_a(M,X)$, and $\tau_{a,\ap}(X)$ is based on samples drawn from the fitted kernel density. 
Table \ref{tabel2} summarizes the performances of $\hat{\psi}_{\text{dml}}$ for sample sizes $n\in \{250,500,1000\}$, where the bias, standard error, and 95\% coverage probability of these estimators averaged across 1000 replications are reported. The debiased machine learning estimator $\hat{\psi}_{\text{dml}}$ exhibits reasonable coverage probabilities under all scenarios with different sample sizes. Moreover, its bias and standard error decrease as the sample size increases. These results corroborate our theoretical findings and demonstrate the advantages of the proposed estimator.

\section{Application}
\label{sec:application}
In this section, we apply the proposed method to data from the 2020 wave of the Health and Retirement Study (HRS). The HRS is a longitudinal panel study conducted by the Institute for Social Research at the University of Michigan. Initiated in 1992 and conducted biennially, it surveys a representative sample of approximately 20000 people aged 50 or older and their spouses in the United States, collecting comprehensive information on health, economic status, and social well-being. 
We examine racial disparities in life satisfaction and the potential mediating pathways underlying them. Prior research by \cite{barger2009relative} and \cite{achdut2024ethnic} has indicated the potential roles of social relationships (e.g., discrimination and loneliness) and socioeconomic status (SES) in accounting for these disparities across racial groups. In this study, we focus on discrimination as the primary mediator of interest. However, as discussed in the Introduction, SES emerges as a potential mediator-outcome confounder that may itself be influenced by race in this case. This analytical challenge is further compounded because SES is an underlying construct that cannot be directly measured. Instead, we would expect the available {variables}, such as income, education level, occupational status, etc., to be imperfect proxies of SES, which help estimate the causal mediation effects under the proposed framework.

In our study, the binary treatment variable $A$ is an indicator of individuals' race (1 for Black or African American, 0 for White or Caucasian). The mediator $M$ indicates the extent to which individuals experience a range of discriminatory behaviors in their daily lives, as measured by the Everyday Discrimination Scale \citep{williams1997racial}. We dichotomize $M$ such that $M = 0$ if the mean score across scale items is greater than 5 (indicating that, on average, the individual experienced specific discriminatory behaviors less than once per year), and $M = 1$ otherwise. 
The outcome variable $Y$ represents the score of a scale, calculated as the average of items from Diener’s measure of Satisfaction with Life \citep{diener1985satisfaction}, with a possible range of 1 to 7. Higher scores indicate higher self-assessed quality of life. The baseline covariates $X$ include age, gender, and marital status; the individual's highest educational degree attained and self-reported health status are treated as measured treatment-induced confounders $L$. {We use mother's and father's years of education as the candidate mediator-inducing proxies $Z$, which reflect one's access to educational and economic resources during their formative years, and use total wealth and total non-housing wealth as the candidate outcome-inducing proxies $W$.} After controlling for observed common causes, the causal pathway from parents' education level to life satisfaction is expected to operate exclusively through the individual's SES, while the association between wealth and discrimination experiences is postulated to be fully explained by SES. These features ensure that the key independence assumptions underlying the use of proxies are plausibly satisfied in our setting.

\begin{table}[h!]
\centering
\caption{Interventional effect estimates (standard errors) and 95\% confidence intervals of race on life satisfaction mediated by discrimination.}
\label{table3}
\resizebox{\textwidth}{!}{
\begin{threeparttable}
\begin{tabular}{cccccc}
\hline
 &  & $\text{IIE}_{\text{mr}}$ & $\text{IIE}_{\text{dml}}$ & $\text{IDE}_{\text{mr}}$ & $\text{IDE}_{\text{dml}}$ \\ \hline
\multirow{2}{*}{$\mathcal{C}_1$} & Estimates (SEs) & -0.029 (1.252) & -0.025 (0.016) & -0.259 (2.092) & -0.314 (0.069) \\
 & 95\% CIs & (-2.482, 2.425) & (-0.057, 0.007) & (-4.360, 3.842) & (-0.450, -0.178) \\
\multirow{2}{*}{$\mathcal{C}_2$} & Estimates (SEs) & -0.024 (0.040) & -0.046 (0.015) & -0.262 (0.097) & -0.287 (0.071) \\
 & 95\% CIs & (-0.103, 0.056) & (-0.077, -0.016) & (-0.452, -0.072) & (-0.427, -0.148) \\
\multirow{2}{*}{$\mathcal{C}_3$} & Estimates (SEs) & -0.024 (0.139) & -0.026 (0.016) & -0.240 (0.857) & -0.288 (0.064) \\
 & 95\% CIs & (-0.297, 0.250) & (-0.058, 0.006) & (-1.920, 1.439) & (-0.415, -0.162) \\
\multirow{2}{*}{$\mathcal{C}_4$} & Estimates (SEs) & -0.010 (0.112) & -0.032 (0.016) & -0.219 (0.310) & -0.295 (0.068) \\
 & 95\% CIs & (-0.230, 0.210) & (-0.064, 0.000) & (-0.826, 0.388) & (-0.428, -0.161) \\ \hline
\end{tabular}
\begin{tablenotes}
    \small
    \item $\mathcal{C}_1$: mother's education ($Z$), total wealth ($W$); $\mathcal{C}_2$: father's education ($Z$), total wealth ($W$); $\mathcal{C}_3$: mother's education ($Z$), total non-housing wealth ($W$); $\mathcal{C}_4$: father's education ($Z$), total non-housing wealth ($W$).
\end{tablenotes}
\end{threeparttable}}
\end{table}

{After excluding subjects with missing data on the mediator and outcome, as well as abnormal individuals under age 50, the final sample consists of 4090 observations, 865 with $A = 1$ and 3225 with $A = 0$.} For parametric estimation, we specify logistic regression models for the treatment $A$ and mediator $M$, as well as linear regression models for the proxy variable $W$ and the measured treatment-induced confounders $L$. Functional forms presented in the simulation studies are employed for the confounding bridge functions $h_a$ and $q_a$. For the debiased machine learning estimation, we adopt the same machine learning techniques and hyperparameter configurations as in the simulation. {As a sensitivity check, we construct four different proxy combinations ($\mathcal{C}_1$-$\mathcal{C}_4$) by taking one candidate from each of the two groups, and carry out the analysis separately. Standard errors for the parametric method are obtained via the nonparametric bootstrap, whereas those for the debiased machine learning method are derived from the EIF.}

Table \ref{table3} summarizes the parametric method-based multiply robust and cross-fitted debiased machine learning estimates of interventional effects in our empirical example. {As can be seen, although the point estimates from both methods are relatively close, the multiply robust estimates $\text{IIE}_{\text{mr}}$ and $\text{IDE}_{\text{mr}}$ suffer from large and unstable standard errors, making the corresponding 95\% confidence intervals non-informative. In contrast, the debiased machine learning estimates $\text{IIE}_{\text{dml}}$ and $\text{IDE}_{\text{dml}}$ remain relatively stable across different proxy combinations. These results corroborate our theoretical findings and suggest a complex data-generating mechanism in the real world. Furthermore, the debiased machine learning estimates for both interventional direct and indirect effects exhibit predominantly negative 95\% confidence intervals, indicating racial disparities in life satisfaction, with a modest yet non-negligible indirect effect through discrimination.} This points to the value of targeted programs and practices aimed at mitigating the harmful effects of discrimination as a critical direction for supporting the well-being of Black and African American older adults.

\section{Discussion}\label{sec:discussion}
In this paper, we propose a novel framework for causal mediation analysis in the presence of unmeasured treatment-induced confounding. We formally articulate the required identification conditions, establish nonparametric identification by leveraging available proxy variables, develop semiparametric theory and propose the corresponding multiply robust estimator. {By taking into account the unmeasured treatment-induced confounding, our approach is particularly suitable for analyzing the effects of treatments that are randomly assigned long before the mediation measurement, or inherent characteristics, such as sex and ethnicity, that are determined at birth but may exert lifelong influences through complex biological and social pathways.}

There are several possible future directions for this line of research. First, longitudinal data have received considerable attention in recent years, and progress in this direction has been made in both mediation analysis \citep{vanderweele2017mediation} and proximal causal inference \citep{ying2023proximal}. It is thus of interest to extend our identification and estimation strategies to a more general longitudinal setting, with time-varying exposures, mediators, and unmeasured treatment-induced confounders. Second, a more challenging scenario arises when both unmeasured pre-treatment confounding and unmeasured treatment-induced confounding exist. Addressing this case effectively is difficult and constitutes an important direction for future research. Third, like most of the work under the proximal causal inference framework, our method relies  on the validity of proxy variables,  which are typically selected based on expert knowledge or subjective judgment. When certain assumptions are violated, methods for sensitivity analysis for the selection and validation of proxy variables may be useful.

\bibliographystyle{apalike}
\bibliography{bibliography}

\end{document}